\documentclass[a4paper,11p]{article}
\pdfoutput=1

\usepackage{jheppub}

\usepackage[font=footnotesize,labelfont=bf]{caption}
\usepackage{graphicx}
\usepackage{subcaption}
\usepackage{bm}
\usepackage{amsmath}
\usepackage{amssymb}
\usepackage{hyperref}
\usepackage{color}
\usepackage{academicons}
\usepackage{xcolor}
\usepackage{fontawesome5}

\newcommand{\dd}{\mathrm{d}}

\newcommand{\CL}{{\tt ${\mathcal C}$osmo${\mathcal L}$attice}~}
\definecolor{orcidlogocol}{rgb}{0.65, 0.807, 0.223}
\newcommand{\orcid}[1]{$\,$\href{https://orcid.org/#1}{\textcolor{orcidlogocol}{\faOrcid}}}

\title{\boldmath Self-resonance during preheating: The case of $\alpha$-attractor models}

\author[a,b]{Daniel del-Corral\orcid{0000-0002-6776-561X}}

\affiliation[a]{Departamento de F\'{\i}sica, Universidade da Beira Interior, Rua Marqu\^{e}s D'\'Avila e Bolama 6200-001 Covilh\~a, Portugal}
\affiliation[b]{Centro de Matem\'atica e Aplica\c{c}\~oes da Universidade da Beira Interior, Rua Marqu\^{e}s D'\'Avila e Bolama 6200-001 Covilh\~a, Portugal}

\emailAdd{corral.martinez@ubi.pt}

\abstract{In this paper, for the first time, we obtain a new class of solutions for the Hill-type differential equations, which emerge in the preheating self-resonance phase of the expanding Universe. We study, in particular, the class of symmetric and asymmetric scalar field potentials coming from the so-called $\alpha$-attractor models of the early Universe cosmology. By making a series expansion of the potential and employing perturbative techniques we reformulate the Mukhanov-Sasaki equation, which captures the dynamics of the curvature perturbation in these models, into a Hill equation. This last includes higher-order terms that were never solved in the literature. Namely, those coming from the cubic and quartic contributions of the scalar field potential. Then, we derive the expressions for the Floquet exponents of the Mukhanov-Sasaki variable. Our analytical results are then compared with numerical computations, showing a good agreement and thus making this method valuable for obtaining theoretical predictions with new observational applications in the contexts of Primordial Black Holes and Scalar-Induced Gravitational Waves.}

\makeatother
\makeatletter
\gdef\@fpheader{}
\makeatother

\begin{document}
\maketitle


\section{Introduction}

The cosmological models known as $\alpha$-attractors \cite{Kallosh:2013daa,Kallosh:2013hoa,Kaiser:2013sna,Kallosh:2015lwa} show an excellent agreement with Planck data \cite{Planck:2018jri} and have universal predictions of observables. For this reason, they have gained a lot of attention, as well as from the point of view of supergravity, where they are most naturally formulated. These models are characterized by the positive, real, dimensionless parameter $\alpha$. In this work, we study the post-inflationary phase of preheating for this class of models, characterized by the inflaton's oscillation around its minimum. This has already been studied in the context of chaotic inflation from the relevant papers of Kofman, Linde, Starobinsky in the 90s \cite{Kofman:1997yn,Kofman:1994rk}. For the $\alpha$-attractors scenario \cite{Iarygina:2018kee,Iarygina:2020dwe,Krajewski:2018moi,Lozanov:2017hjm}, these oscillations produce a huge amplification of perturbations for small values of $\alpha$. This effect is called self-resonance and can be so strong that the non-linearity regime is reached in $\mathcal{O}(1)$ e-folds after the end of inflation. See \ref{sec:appendix} for details. The $\alpha$-attractor models have been also studied in the context of parametric resonance. However, while many works \cite{Iarygina:2018kee,Iarygina:2020dwe,Krajewski:2018moi,Lozanov:2017hjm,Jedamzik:2010dq,Martin:2019nuw,Sfakianakis:2018lzf,Ballesteros:2024hhq,Martin:2020fgl,Shafi:2024jig,Mahbub:2023faw,Sang:2020kpd,Zhang:2023hjk,Child:2013ria} focus on a quadratic minimum or rely on numerical simulations, we instead expand the potential up to fourth order and develop an analytical method to explain the huge amplification produced by the self-resonance effects. Specifically, and contrary to the usual procedure, our focus lies on the study of the amplification of curvature perturbations through the Mukhanov-Sasaki (MS) equation (see \ref{sec:appendix-MS}). We consider these perturbations to be more relevant from the observational point of view and thus we find it necessary to develop the analytical tools necessary to explain its amplification during preheating. We perform a series expansion of the potential and posterior transformation of the MS equation into a Hill equation \cite{drazin_1992,nayfeh2008nonlinear}, which gives us the Floquet's exponents governing the amplification of the perturbations. This Hill equation, that frequently emerge in the context of preheating, has never been solved including the terms coming from both the cubic and quartic contributions of the expansion of the potential. The idea of including this terms in the expansion of the inflationary potential comes from the fact that, as $\alpha$ decreases, they become relevant, and thus self-resonance comes into play more robustly, as we will shortly see. There are two types of $\alpha$-attractors, the so-called T- and E-models, whose potentials are given, respectively, by \cite{Kallosh:2015lwa}
\begin{subequations}\label{eq:potentials}
    \begin{equation}\label{eq:potential-T}
         V_T(\phi)=3\alpha M^2\tanh^2\left(\frac{\phi}{\sqrt{6\alpha}}\right),
    \end{equation}\begin{equation}\label{eq:potential-E}
         V_E(\phi)=\frac{3\alpha M^2}{4}\left(1-e^{-\sqrt{\frac{2}{3\alpha}}\phi}\right)^2.
     \end{equation}
 \end{subequations}
Here, $M$ is the inflaton mass, to be fixed by CMB normalization, $\phi$ is the scalar field and $\alpha$ is the parameter characterizing the models, with some constraints\footnote{In \cite{Iacconi:2023mnw} it is given a lower bound for the parameter $\alpha$, $\log_{10}\alpha=-4.2_{-8.6}^{+5.4}$ (95\% C.L.), based on CMB constraints. We consider values of $\alpha$ above this lower bound for this work.}. Expanding the potentials around $\phi=0$ we get
\begin{subequations}
\begin{equation}\label{eq:expansion-T}
        V_T(\phi)=\frac{M^2}{2}\phi^2+\frac{\lambda_T}{4}\phi^4+\dots,
    \end{equation}
    \begin{equation}\label{eq:expanson-E}
        V_E(\phi)=\frac{M^2}{2}\phi^2+\frac{\lambda_3}{3}\phi^3+\frac{\lambda_E}{4}\phi^4+\dots
,
    \end{equation}
\end{subequations}
where the coefficients of the cubic and quartic terms, responsible for the anharmonic behavior of the field, are given in terms of the parameter $\alpha$ by
\begin{equation}\label{eq:parameters-potential}
    \lambda_T=-\frac{2M^2}{9\alpha},\qquad\lambda_3=-M^2\left(\frac{3}{2\alpha}\right)^{1/2},\qquad\lambda_E=\frac{7M^2}{9\alpha}.
\end{equation}
This, as we pointed above, indicates that decreasing the parameter $\alpha$ enhances the importance of the cubic and quartic terms, even for relatively small values of $\phi$. The results obtained here can be extrapolated to any potential whose Taylor expansion is given by a quadratic part + higher order terms. Furthermore, the implications of our study extend to various phenomena. If the amplification from self-resonance reaches significant levels, perturbations could collapse and form Primordial Black Holes (PBHs) \cite{Zeldovich:1967lct,Hawking:1971ei,Carr:1974nx,Carr:1975qj} or Scalar-Induced Gravitational Waves (SIGWs) \cite{Maggiore:2007,Guzzetti:2016,Domenech:2021,Baumann:2007zm,Ananda:2006af}. We leave this as a future project \cite{draft-alpha-PBH,draft-alpha-GWs}. 

The contents of this work are organized as follows. We begin in Sec.~\ref{sec:self-resonance} by studying the effect of the self-resonance on a generic potential and computing the Floquet's exponents. Then in Sec.~\ref{sec:characterization} we show specifically for a T- and E-model the amplification of curvature perturbations and compare the results with a numerical computation. Sec.~\ref{sec:contrast} focuses on the importance of considering higher order terms in the expansion of the potential by comparing with the parabola approximation and finally in Sec.~\ref{sec:applications} we give some applications where this work can be useful. Conclusions are given in Sec.~\ref{sec:conclusions}, \ref{sec:appendix-MS} shows the derivation of the MS-equation, and \ref{sec:appendix} shows the time until the non-linearity regime is reached as well as some comparisons with lattice simulations.


\section{Self-resonance for a generic potential} \label{sec:self-resonance}

To obtain a general result for both T- and E-models, we will study self-resonance in potentials of the form
\begin{equation}\label{eq:potential-generic}
    V(\phi)=\frac{M^2}{2}\phi^2+\frac{\lambda_3}{3}\phi^3+\frac{\lambda}{4}\phi^4.
\end{equation}
The analysis is general and valid for every potential that can be written in this form, regardless of the presence of higher-order terms such as $\sim\phi^5$, $\sim\phi^6$, \dots


\subsection{Perturbation theory for anharmonic oscillators}\label{sec:perturbation-theory}

\begin{figure}
     \centering
     \begin{subfigure}[b]{0.4\textwidth}
         \centering         \includegraphics[width=\textwidth]{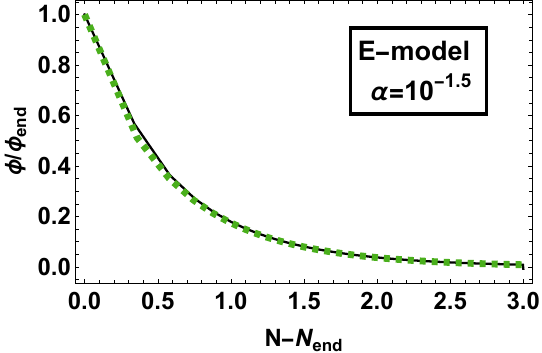}
         \caption{}
         \label{fig:amplitudeE15}
     \end{subfigure}
     \begin{subfigure}[b]{0.4\textwidth}
         \centering         \includegraphics[width=\textwidth]{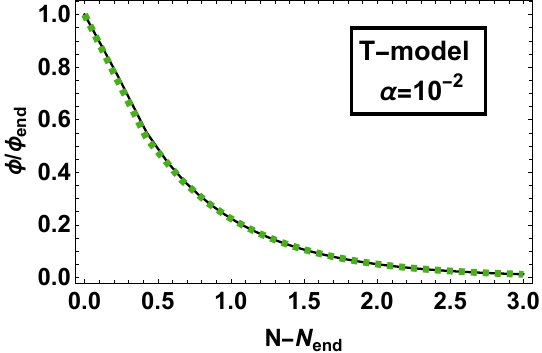}
         \caption{}
         \label{fig:amplitudeT2}
     \end{subfigure}
     \begin{subfigure}[b]{0.4\textwidth}
         \centering         \includegraphics[width=\textwidth]{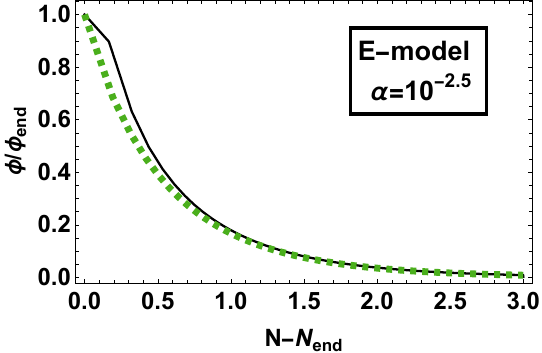}
         \caption{}
         \label{fig:amplitudeE25}
     \end{subfigure}
     \begin{subfigure}[b]{0.4\textwidth}
         \centering         \includegraphics[width=\textwidth]{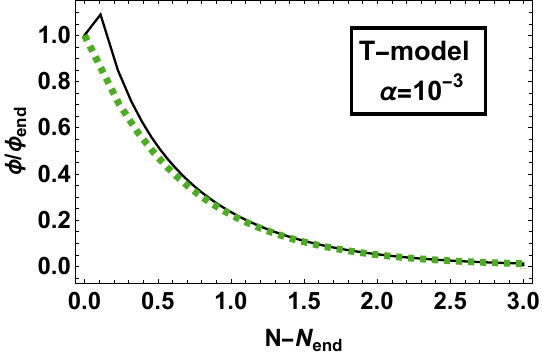}
         \caption{}
         \label{fig:amplitudeT3}
     \end{subfigure}
        \caption{Amplitude of the field for E- and T-models with different values of alpha. In each case, the perturbative expansion \eqref{eq:final-form} is shown in dotted green and the numerical solution is in continuous black.}
        \label{fig:amplitudes}
\end{figure}

The background dynamics of the inflaton field $\phi$ are governed by the equation of motion of the field and Friedmann equations \cite{Baumann:2009ds,Mukhanov:1990me}
\begin{subequations}
\begin{equation}\label{KG}
\ddot{\phi}+3H\dot{\phi}+\frac{\dd V(\phi)}{\dd\phi} =0\,,
\end{equation}
\begin{equation}\label{FR}
    H^2=\frac{1}{3M_p}\left(\frac{\dot\phi^2}{2}+V_{T,E}(\phi)\right),
\end{equation}
\end{subequations}
respectively, where a dot means derivation with respect to cosmic time, $H=\dot a/a$ is the Hubble rate, $a$ is the scale factor of the universe, and $M_p$ is the Planck mass. After inflation, the friction term in \eqref{KG} is sub-dominant and thus we can neglect it. Later, we will revisit the issue of considering the Hubble expansion. Under this assumption, eqn.~\eqref{KG} can be solved using perturbation theory \cite{drazin_1992,nayfeh2008nonlinear}. Among all the perturbative methods available, we choose to work with the Lindsted-Poincaré Method\footnote{Other methods include the Time Transformation Method \cite{BURTON1983543}, the Modified Lindsted-Poincaré Method \cite{CHEUNG1991367}, the Multiple Scales Method \cite{nayfeh2008nonlinear}, or some methods based on Jacobian elliptic functions \cite{Kovacic-2016}. We however restrict to the Lindsted-Poincaré method due to its simplicity and good results.}. First, we change the differential equation \eqref{KG} to a time domain where the solution is simple harmonic and of period $2\pi$. This is done by considering $\tau=\omega t$, where $\omega$ is the real frequency of oscillation of $\phi$. Due to the presence of higher order terms in the expansion of the potential, the frequency is not simply given by $\omega^2=M^2$. Instead, we parametrize it as $\omega^2=M^2(1-\beta^2)$, where $\beta$ is a small, real parameter that carries the information about the departure of the system from quadratic behavior. We are assuming a negative sign in the parametrization of $\omega^2$ due to the negative sign in the parameters $\lambda_T$ and $\lambda_E$, which makes the potential to be, in general, wider than quadratic, thus decreasing the frequency of oscillations of the field. Next, the field is expanded into powers of the same small parameter $\beta$ as follows
\begin{equation}\label{eq:field-expansion}
    \phi=\beta\phi_1+\beta^2\phi_2+\beta^3\phi_3+\dots,
\end{equation}
with initial conditions $\phi_i(0)=\Phi_i$ and $\phi_i'(0)=0$. If we now substitute this expansion into \eqref{KG} and use \eqref{eq:potential-generic}, we find that at the lowest order, $\mathcal{O}(\beta)$, the system behaves as a simple harmonic oscillator
\begin{equation}
    \phi_1=\Phi_1\cos\tau.
\end{equation}
Now, going to the next order, $\mathcal{O}(\beta^2)$, we observe that $\phi_2$ is a combination of the zeroth, first and second harmonics, that is
\begin{equation}
    \phi_2=\frac{\lambda_3\Phi_1^2}{6M^2}\left(\cos(2\tau)-3+2\cos(\tau)\right)+\Phi_2 \cos(\tau).
\end{equation}
Following the approach of the time transformation method \cite{CHEUNG1991367,BURTON1983543,nayfeh2008nonlinear}, the initial conditions for $\phi_i$ ($i>1$) can be chosen arbitrarily, as long as the main initial condition $\phi(0)=\phi_0$ holds. Thus, we can choose the amplitude $\Phi_2$ to cancel out the first odd harmonic. In this sense, the second-order solution contains only even harmonics, which enable us to obtain a more accurate solution \cite{CHEUNG1991367}. Going now to the next order, $\mathcal{O}(\beta^3)$, we encounter that in the differential equation, there are some terms proportional to $\cos(\tau)$ depending on $\lambda_3$ and $\lambda$, which produces a secular behavior \cite{drazin_1992}. This is because some of the driving terms in the differential equation have exactly the fundamental frequency of the system. To avoid this, we choose the following value for the amplitude $\Phi_1$
\begin{equation}
    \Phi_1=\frac{2M}{\sqrt{3\left|\lambda-\frac{10\lambda_3^2}{9M^2}\right|}},
\end{equation}
so that now the solution $\phi_3$ does not contain secular terms. Choosing again $\Phi_3$ so that it cancels out the first odd harmonic, we obtain a solution for $\phi_3$ in terms of just the third odd harmonic as
\begin{equation}
    \phi_3=\Phi_1^3\left(\frac{\lambda}{32M^2}+\frac{\lambda_3^2}{48M^4}\right)\cos(3\tau).
\end{equation}
We find it enough for our purposes to work up to order $\mathcal{O}(\beta^2)$, for which the solution is
\begin{equation}\label{eq:final-form}
    \phi=\beta\Phi_1\cos(\tau)+\beta^2\frac{\lambda_3\Phi_1^2}{6M^2}(\cos(2\tau)-3).
\end{equation}
For the initial condition $\phi(0)=\phi_0$ to hold, we must solve for $\beta$ at $\tau=0$ which, as previously said, is the correction to the frequency, which we defined as $\omega^2=M^2(1-\beta^2)$. Once solved it will be given in terms of the anharmonic terms of the potential, namely $\lambda_3$ and $\lambda$, and in terms of the amplitude of the field $\phi_0$. The larger the amplitude, the larger the correction to the frequency since the anharmonic terms became more dominant. For a small amplitude, we have $\omega^2\simeq M^2$ and thus we recover the quadratic case (parabolic potential). In this sense, we see that including the friction term, and thus the decay of the field, the correction term $\beta$ decreases and the system moves towards the quadratic approximation of the potential. Therefore, one can approximate this situation by choosing a $\beta$ that decays at the same rate as the field. That is
\begin{equation}
    \beta(\tau)=\beta_0\left(\frac{a_0}{a}\right)^{3/2},
\end{equation}
where $\beta_0$ is obtained from \eqref{eq:final-form} at $\tau=0$, considered as the end of inflation, and $a$ is the scale factor of the universe, with $a_0=a(0)$. To simplify notation we omit the argument of $\beta$ and neglect its derivatives since it is assumed to vary slowly. Fig.~\ref{fig:amplitudes} shows the decay of the normalized amplitude of the field for the T- and E-models with different values of $\alpha$ using the perturbative expansion \eqref{eq:final-form}. For comparison, it has also been included the amplitude of the numerical solution of the equation of motion of the field, eqn,~\eqref{KG}, which has been solved self-consistently with the Friedmann equation \eqref{FR}. As one can observe, the agreement is good, but as we decrease the values of $\alpha$ the analytic and numerical solutions deviate during the first e-fold of preheating. This is due to the lack of higher order terms in the expansion \eqref{eq:final-form}, which contribute more as $\alpha$ decreases. However, as Fig.~\ref{fig:curvatures} shows, working up to second order in $\beta$ is enough for our purposes.

The term $\beta^2\lambda_3\Phi_1^2/2M^2$ in the solution for $\phi_2$ is often called the drift or steady streaming term \cite{nayfeh2008nonlinear}, characteristic of systems with even non-linearities ($\lambda_3\phi^2$). It carries information about the asymmetry of the potential, since as we observe it is directly proportional to $\lambda_3$. Thus we clearly see that the asymmetry of the potential is transferred to the solution of the field, which will have a significant impact on the self-resonance effect, as we will see. For symmetric potentials ($\lambda_3=0\not=\lambda$), this self-resonance effect is not as pronounced as in the asymmetric case, since the solution is given naturally in terms of odd harmonics, which is closer to the quadratic case.


\subsection{Floquet theory for perturbations}\label{sec:floquet-pert}

The evolution of the scalar perturbations is controlled by the MS variable $v_{\bm k}$ \cite{Mukhanov:1990me,Baumann:2009ds},  where a suffix $\bm k$ means we are working in Fourier space. The MS equation expressed in the time variable $\tau$ can be written as 
\begin{equation}\label{eq:hill-eq}
    \frac{\dd^2\tilde{v}_{\bm k}}{\dd\tau^2} +\left(A_k+2q_1\cos(\tau)+2p_1\sin(\tau)+2q_2\cos(2\tau)+2p_2\sin(2\tau)+h.h.\right)\tilde{v}_{\bm k}=0,
\end{equation}
where $h.h.$ stands for higher harmonics\footnote{We do not consider higher harmonics since doing so would introduce a higher order in $\beta$ when calculating the Floquet exponent \cite{Amin:2010xe}}, and the coefficients $A_k$, $q_i$ and $p_i$, as well as the derivation of the MS-equation, are specified in \ref{sec:appendix-MS}.
Eqn.~\eqref{eq:hill-eq} is known as a Hill equation, a generalization of a Mathieu equation \cite{Hertzberg:2014iza,Hertzberg:2014jza,Jedamzik:2010dq,Kofman:1997yn}. Depending on the values of the parameters $A_k$, $q_i$, and $p_i$ the physical modes $(k/a)^{-1}$ experience instability or stability as they evolve. Following Floquet's theorem, the re-scaled MS variable evolves as $\tilde{v}_{\bm k}\sim \exp{\bigl[\int\mu_k(\tau)\dd \tau\bigr]}$, where $\mu_k$ are the so-called \textit{Floquet exponents}. For $\Re(\mu_k)>0$ we have exponential growth and the mode is said to be unstable. The Floquet exponents of the Hill equation are generally unknown but can be computed using the harmonic balance method \cite{nayfeh2008nonlinear}, where the MS variable $\tilde{v}_{\bm k}$ is expanded into harmonics as
\begin{equation}\label{eq:harmonic-expansion}
    \tilde{v}_{\bm{k}}=\sum_{n=0}^{\infty}e^{in\tau}C_n(\tau).
\end{equation}
By substituting this into \eqref{eq:hill-eq} and expressing the trigonometric functions as complex exponential functions, the following recurrence relation is obtained
\begin{equation}\label{eq:recurrence}
    C_n'=\frac{i}{2n}\left[(A_k-n^2)C_n+q_1(C_{n+1}+C_{n-1})+ip_1(C_{n+1}-C_{n-1})+q_2(C_{n+2}+C_{n-2})+ip_2(C_{n+2}-C_{n-2})\right],
\end{equation}
where we are neglecting the $C_n''$ since we assume that $C_n$ are slowly varying. The first instability band emerges from examining the fundamental frequencies $n=\pm1$ in \eqref{eq:recurrence}. Tipically, to order $\beta^2$, the coefficients $C_{\pm1}$ are coupled just one to each other. However, in our case, the presence of $q_1$ and $p_1$, along with $q_1$ and $p_2$ being of order $\beta$, the coefficients $C_{\pm1}$ are also being coupled to $C_0$, $C_{\pm2}$ and $C_{\pm3}$ to order $\beta^2$. This additional coupling complicates the computation of the Floquet exponents. Nonetheless, we have found that it suffices for our purposes to consider the fundamental frequencies $n=\pm1$ and their corresponding couplings with the nearest harmonics $n=0,2$ and $n=-2,0$, respectively, up to order $\beta^2$. By doing this, we derive the following differential matrix equation
\begin{equation}\label{eq:matrix-eq}
\begin{pmatrix}
        C_1' \\ C_{-1}'
    \end{pmatrix}\simeq\frac{i}2
    \begin{pmatrix}
        A_k-1-\frac{q_1^2+p_1^2}{A_k}-\frac{q_1^2+p_1^2}{A_k-4} & q_2-ip_2-\frac{(q_1-ip_1)^2}{A_k} \\
        -q_2-ip_2-\frac{(q_1+ip_1)^2}{A_k} & 1-A_k+\frac{q_1^2+p_1^2}{A_k}+\frac{q_1^2+p_1^2}{A_k-4}
    \end{pmatrix}
    \begin{pmatrix}
        C_1 \\ C_{-1}
    \end{pmatrix}.
\end{equation}
Finally, remembering Floquet's theorem, the re-scaled MS variable evolves as $\tilde{v}_{\bm k}\simeq \exp{\bigl[\int\mu_k(\tau) \dd \tau\bigr]}$ \cite{Jedamzik:2010dq}, where $\mu_k(\tau)$ are now given by the eigenvalues of the matrix in \eqref{eq:matrix-eq}. Up to order $\beta^2$, these are given by:
\begin{equation}\label{eq:floquet-exponent-general}
\begin{split}
    \mu_k=\frac12\Biggl[q_2^2+p_2^2-(A_k-1)^2+2q_1^2\Big(\frac{A_k-1}{A_k}+&\frac{A_k-1}{A_k-4}-\frac{q_1^2}{2(A_k-4)^2}-\\&-\frac{q_1^2}{A_k(A_k-4)}\Bigr)+2p_1^2\left(\frac{A_k-1}{A_k}+\frac{A_k-1}{A_k-4}\right)-\frac{2q_1^2q_2}{A_k}-\frac{4p_1p_2q_1}{A_k}\Biggr]^{1/2}.
\end{split}
\end{equation}
We can relate the MS variable to the comoving curvature perturbation $\mathcal{R}_{\bm k}$ by \begin{equation}\label{eq:curvature-definition}
\mathcal{R}_{\bm k}=\frac{v_{\bm k}}{\sqrt{2\epsilon_H}a}=\frac{\tilde{v}_{\bm k}}{\sqrt{2\epsilon_H a^3}},
\end{equation}
where $\epsilon_H=-\dot{H}/H^2$ is the Hubble slow-roll parameter in cosmic time. Eqns.~\eqref{eq:floquet-exponent-general} and \eqref{eq:curvature-definition} give us the time-evolution of the curvature perturbation as a function of the anharmonic coefficients of the potential. This allows us to efficiently give an estimation of the amplification of perturbations along with the position of the peak in terms of the comoving wavenumber $k$. This has some applications as we show in Sec.~\ref{sec:applications}. The Floquet's exponents of the curvature perturbations are, to our knowledge, shown for the first time for an inflationary potential presenting both cubic and quartic terms in its expansion. This implies a great advantage from the numerical point of view since traditionally the codes designed to compute Floquet's exponents must deal with highly oscillatory functions that increase the computational time and become rather inefficient. From \eqref{eq:curvature-definition} we see that when the amplification of $\tilde{v}_{\bm k}$ exceeds $a^{3/2}$, the curvature perturbation will grow. Typically, for potentials close to parabolic, this amplification is as high as $a^{3/2}$, causing $\mathcal{R}_{\bm k}$ to remain approximately constant for the modes inside the resonance band \cite{Martin:2019nuw}. However, as we demonstrate in the next section, deviating from quadratic behavior alters the scenario, producing a fast amplification in a specific range of modes.


\section{Characterization of self-resonance}\label{sec:characterization}

Here we present the specific Floquet's exponents for both T- and E-models and compare them with numerical results. These findings can be generalized to any potential whose expansion around $\phi=0$ matches either the expansion of a T- or an E-model.
\begin{figure}
     \centering
     \begin{subfigure}[b]{0.45\textwidth}
         \centering         \includegraphics[width=\textwidth]{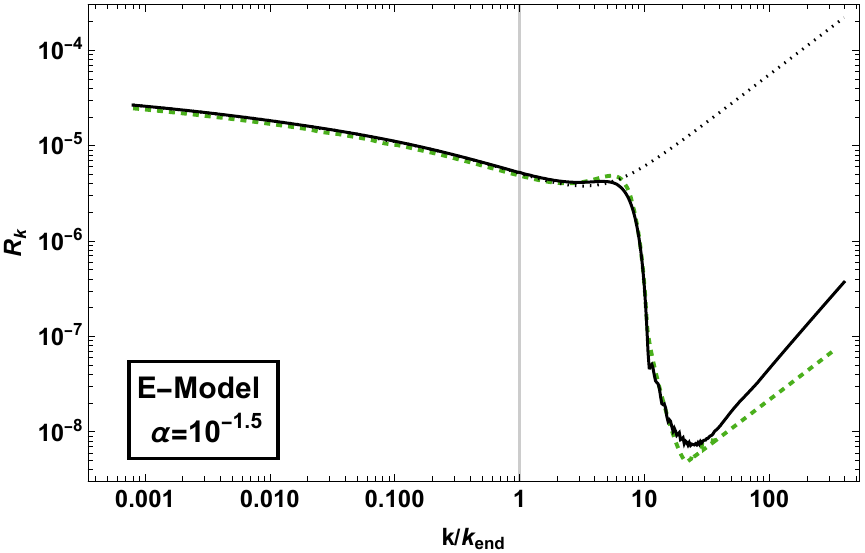}
         \caption{}
         \label{fig:curvature-E15}
     \end{subfigure}
     \begin{subfigure}[b]{0.45\textwidth}
         \centering         \includegraphics[width=\textwidth]{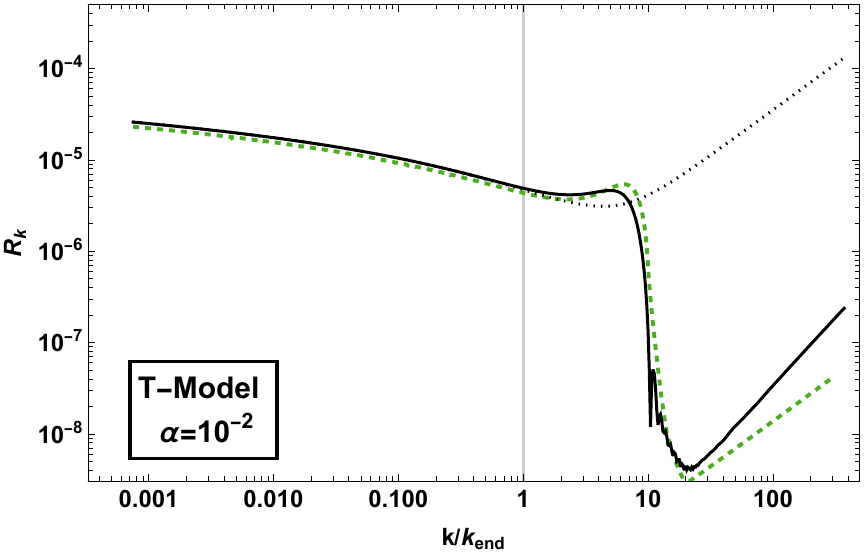}
         \caption{}
         \label{fig:curvature-T2}
     \end{subfigure}
     \begin{subfigure}[b]{0.45\textwidth}
         \centering         \includegraphics[width=\textwidth]{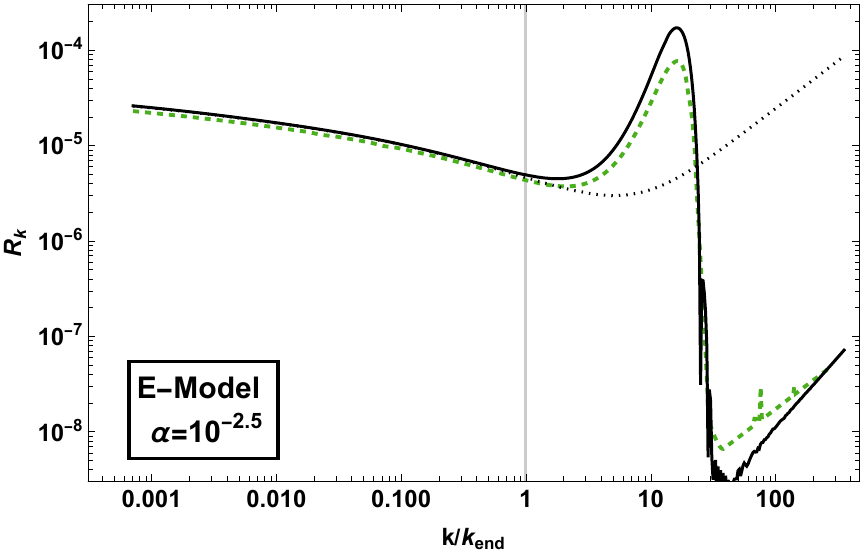}
         \caption{}
         \label{fig:curvature-E25}
     \end{subfigure}
     \begin{subfigure}[b]{0.45\textwidth}
         \centering         \includegraphics[width=\textwidth]{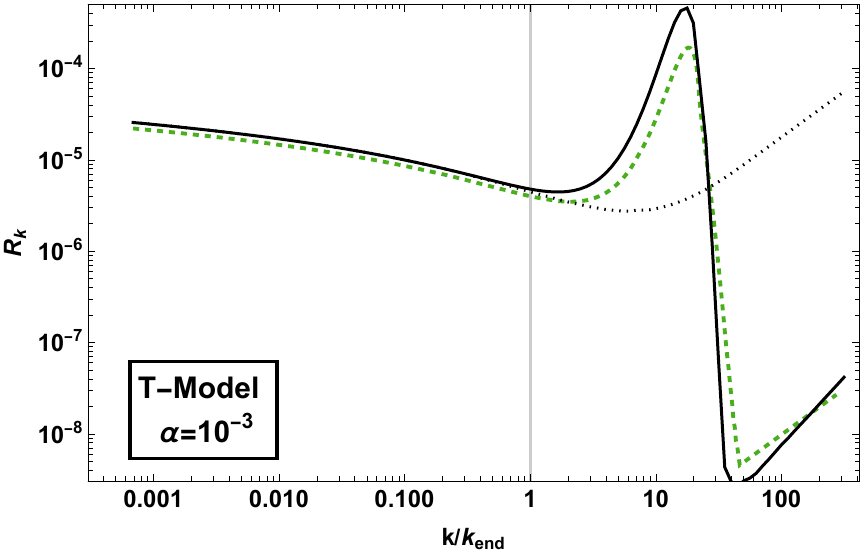}
         \caption{}
         \label{fig:curvature-T3}
     \end{subfigure}
        \caption{Curvature perturbations for \textbf{(a)} E-Model $\alpha=10^{-1.5}$, \textbf{(b)} T-Model $\alpha=10^{-2}$, \textbf{(c)} E-Model $\alpha=10^{-2.5}$ and \textbf{(d)} T-Model $\alpha=10^{-3}$ evaluated 5 e-folds after the end of inflation. Numerical computation is shown in continuous black and Floquet theory as described in this work in dashed green. The vertical grey line marks the scale that exits the horizon at the end of inflation, $k_{\text{end}}$. For comparison, black dotted curves show the numerical computation of curvature perturbations at the end of inflation.}
        \label{fig:curvatures}
\end{figure}


\subsection{T-model}

\begin{figure}
     \centering
     \begin{subfigure}[b]{0.45\textwidth}
         \centering         \includegraphics[width=\textwidth]{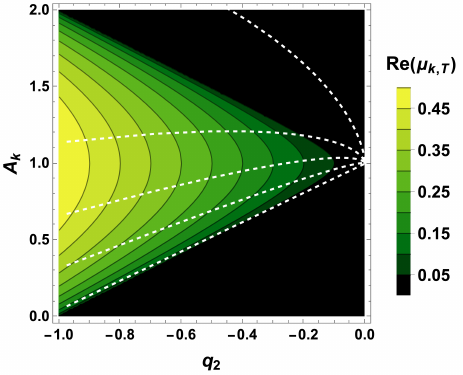}
         \caption{}
         \label{fig:chart-trajectories-T}
     \end{subfigure}
     \begin{subfigure}[b]{0.45\textwidth}
         \centering         \includegraphics[width=\textwidth]{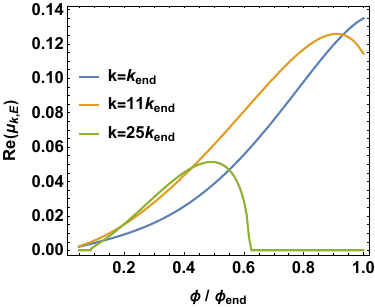}
         \caption{}
         \label{fig:floquet-evolution-E}
     \end{subfigure}
     \caption{\textbf{(a)} Floquet chart for a T-model with $p_2=0$ (just for graphical purposes) and $\alpha=10^{-2.5}$. White dashed lines correspond to the evolution of the modes in the $(q_2,A_k)$ plane and are, from bottom to top: 1, 10, 15, 20, 30, in units of $k_{\text{end}}$. \textbf{(b)} Evolution of $\Re(\mu_{k,E})$ for an E-model with $\alpha=10^{-2.5}$ as a function of the amplitude of the field and for different values of $k$.}
\end{figure}

In this case we have that $\lambda_3=0$ and thus $q_1=p_1=0$, which greatly simplifies the expression for the Floquet's exponent
\begin{equation}\label{eq:floquet-T}
    \mu_{k}^{(T)}=\frac12\sqrt{q_2^2+p_2^2-(A_k-1)^2}.
\end{equation}
To check the validity of this approximation, we have computed the curvature perturbations numerically, using \eqref{eq:MS-tau} with Bunch-Davies initial conditions and \eqref{eq:curvature-definition}. Then, we have used the Floquet theory derived here, where the scale factor, needed to parametrize the decay of the field, is obtained by self-consistently solving the Friedmann equation \eqref{FR}, together with the equation of motion of the field \eqref{KG}. This is important, since considering that the time dependence of the scale factor follows a matter-dominated power-law, that is, $a\sim t^{2/3}$, is incorrect during the first e-folds of preheating, which is precisely where this works mainly focus. Both methods are evaluated 5 e-folds after the end of inflation. Results are displayed in Figs.~\ref{fig:curvature-T2} and \ref{fig:curvature-T3} for $\alpha=10^{-2}$ and $\alpha=10^{-3}$ and it shows a good agreement between both methods. Going to smaller values of $\alpha$ implies entering into the non-linear regime, where perturbations backreact and spoil the linear evolution. See \ref{sec:appendix} for further details and a comparison with previous estimations. The peaks to the right of the highest peak in Fig.~\ref{fig:curvatures} correspond to the amplification in higher resonance bands. One could study these bands by considering the frequencies $n=\pm2,\pm3\dots$ and their corresponding couplings. However, to do so, one has to increase each time the order in $\epsilon$, increasing the number of harmonics that come into play. We find it sufficient for our purposes to focus on the first resonance band, as it is where the amplification is most pronounced. Finally, in Fig.~\ref{fig:chart-trajectories-T} we display $\Re(\mu_{k}^{(T)})$ for $\alpha=10^{-2.5}$ as a function of $q_2$ and $A_k$, as well as the trajectories of some modes in the $(q_2,A_k)$ plane. We remark that in Fig.~\ref{fig:chart-trajectories-T}, $p_2=0$ and thus the $\mu_{k}^{(T)}$ shown does not correspond to the real time-evolution. However, it is enough to understand why the peak is produced for scales around $k\simeq10\,k_\text{end}$. Those scales cross the $(q_2,A_k)$ plane throughout the maximum of $\mu_{k}^{(T)}$. Also, it is possible to give an analytical estimation of the position of the peak from the Floquet's exponent \eqref{eq:floquet-T}. For $k<k_{\text{end}}$, all scales show the same evolution in the $(q_2,A_k)$ plane. This is the reason why all of them are amplified in the same amount.


\subsection{E-model}

Now we have that $\lambda_3\neq0$ and thus the expression for the Floquet exponent is the full one given in \eqref{eq:floquet-exponent-general}, we will call it $\mu_{k}^{(E)}$. Figs.~\ref{fig:curvature-E15} and \ref{fig:curvature-E25} display the curvature perturbations computed using the same method as in Figs.~\ref{fig:curvature-T2} and \ref{fig:curvature-T3} (see previous section), this time for the E-model and for $\alpha=10^{-1.5}$ and $\alpha=10^{-2.5}$. Despite a numerical factor of $\mathcal{O}(1)$ of discrepancy between the numerics and Floquet theory, we find it enough to explain at which scales occur the maximum amplification, which could be also obtained analytically from the full expression \eqref{eq:floquet-exponent-general}. The discrepancy between numerics and the analytical result could be because we are neglecting some higher harmonics, which in the case of the E-model could contribute more significantly than in the T-model, as well as more couplings between them. We have observed analytically and also numerically that the amplification of $\mathcal{R}_{\bm k}$ is higher for the E-model compared to a T-model with the same value of $\alpha$. This is due to the asymmetry of the potential since the perturbative solution for the E-model has even harmonics and, particularly, a drift term that appears already at order $\beta^2$. This fact makes the Floquet's exponent \eqref{eq:floquet-exponent-general} have more contributions that enhance the amplification of the perturbations. Additionally, we have also observed that the position of the peak of maximum amplification is at higher $k$ for the E-model, compared to a T-model with the same $\alpha$. Now, regarding the time evolution of $\mu_{k}^{(E)}$ for each $k$-mode, a plot like Fig.~\ref{fig:chart-trajectories-T} would not be sufficiently illustrative, given that $\mu_{k}^{(E)}$ depends on 5 parameters. This makes it challenging to understand why the peak is around $k\sim11\,k_{\text{end}}$. Instead, Fig.~\ref{fig:floquet-evolution-E} shows $\Re(\mu_{k}^{(E)})$ for $\alpha=10^{-2.5}$ as a function of the amplitude of the field for different values of $k$, expressed in units of $k_{\text{end}}$. Here, we observe that for scales around $k\simeq11\,k_{\text{end}}$, the Floquet exponent exceeds that of other $k$-modes, resulting in higher amplification.


\section{Contrast with quadratic approximation}\label{sec:contrast}

\begin{figure}
     \centering
     \begin{subfigure}[b]{0.49\textwidth}
         \centering         \includegraphics[width=\textwidth]{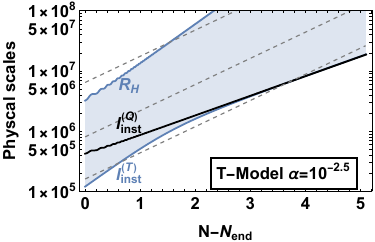}
         \caption{}
         \label{fig:scales-T}
     \end{subfigure}
     \hfill
     \begin{subfigure}[b]{0.49\textwidth}
         \centering         \includegraphics[width=\textwidth]{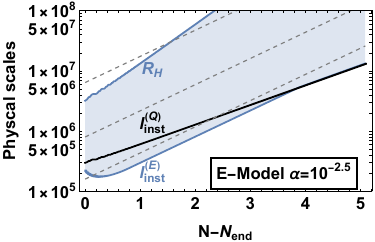}
         \caption{}
         \label{fig:scales-E}
     \end{subfigure}
     \caption{Instability band (blue shaded area) and instability scales, $l_{\text{inst}}$, for the quadratic case and for a \textbf{(a)} T-model and \textbf{(b)} E-model. Also plotted are the Hubble radius $R_H$ and the evolution of some physical modes $(\frac ka)^{-1}$ (gray dashed lines). The horizontal axis represents the number of e-folds elapsed from the end of inflation.}
     \label{fig:scales}
\end{figure}

During preheating, the potential is usually approximated by a parabola ($\lambda_3=\lambda=0$), where the solution for the field is just given by $\phi\simeq\phi_{\text{end}}\left(\frac{a_{\text{end}}}{a}\right)^{3/2}\cos(\tau)$ \cite{Kofman:1997yn,Jedamzik:2010dq,Martin:2020fgl,Martin:2019nuw,Ballesteros:2024hhq}. Considering this and keeping just the dominant terms, \eqref{eq:MS-tau} is easily reformulated into the following Mathieu equation
\begin{equation}
    \tilde{v}_{\bm k}''+\bigl(A_k+2p_2\sin(2z)\bigr)\tilde{v}_{\bm k}=0,
\end{equation}
where now $\tau=Mt$ and
\begin{equation}
    A_k=\frac{k^2}{a^2M^2}+1,\qquad p_2=-\sqrt{\frac32}\phi_{\text{end}}\left(\frac{a_{\text{end}}}{a}\right)^{3/2}.
\end{equation}
Following the steps of Sec.~\ref{sec:floquet-pert}, the Floquet exponent is given by
\begin{equation}\label{eq:floquet-quad}
    \mu_k^{(Q)}=\frac12\sqrt{p_2^2-(A_k-1)^2}.
\end{equation}
\begin{figure}
     \centering
     \begin{subfigure}[b]{0.49\textwidth}
         \centering         \includegraphics[width=\textwidth]{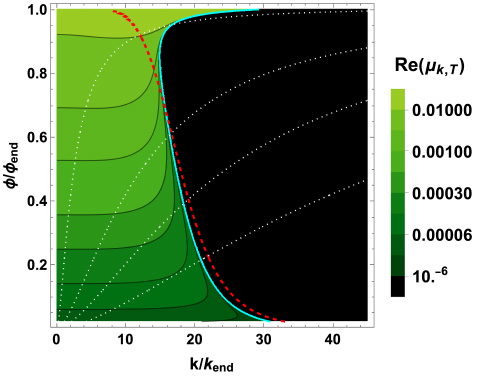}
         \caption{}
         \label{fig:compare-T}
     \end{subfigure}
     \begin{subfigure}[b]{0.49\textwidth}
         \centering         \includegraphics[width=\textwidth]{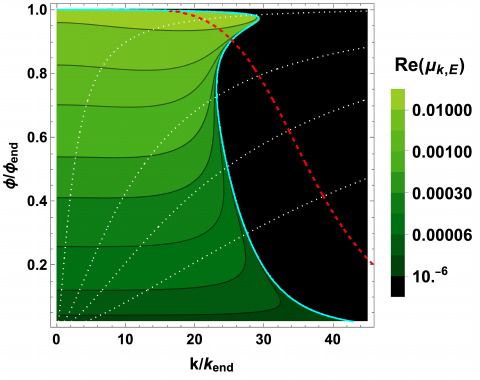}
         \caption{}
         \label{fig:compare-E}
     \end{subfigure}
        \caption{Floquet exponents for \textbf{(a)} T-model and \textbf{(b)} E-model with $\alpha=10^{-2.5}$ as a function of physical wavenumbers and field amplitudes. Red dashed lines represent the instability scale \eqref{eq:spatial-scale} and continuous cyan lines the corresponding instability scales for the full expansion \eqref{eq:potential-generic}. For the T-model, this last is given by \eqref{eq:instab-scale-T}, whereas for the E-model is obtained numerically. White dotted lines mark the evolution of the physical $k$-modes. The black region represents the points where $\Re(\mu_k^{(T,E)})=0$.}
        \label{fig:compare-quadratic}
\end{figure}
Let us now introduce the concept of \textit{instability scale}, denoted as $l_{\text{inst}}$, which represents the spatial scale above which $\tilde{v}_k$ experiences growth. For this growth to occur, the Floquet exponent must be real and positive. Imposing this in \eqref{eq:floquet-quad} and solving for $\left(\frac ka\right)^{-1}\sim l_{\text{inst}}$, one obtains
\begin{equation}\label{eq:spatial-scale}
l_{\text{inst}}^{(Q)}=\frac1{M}\left(\frac2{3\phi_{\text{end}}^2}\left(\frac{a}{a_{\text{end}}}\right)^{3}\right)^{1/4}\simeq\frac1{\sqrt{3HM}}.
\end{equation}
This is the standard preheating scenario. However, this approximation is not accurate for some potentials, even if they behave as quadratic at first order, such as the case of $\alpha$-attractors with $\alpha\ll1$. Fig.~\ref{fig:compare-quadratic} corroborates this fact. It shows the Floquet exponent for both the T- and E-models and compares, in each case, $l_{\text{inst}}^{(Q)}$ (dashed red) with the corresponding spatial scale $l_{\text{inst}}^{(T,E)}$ (cyan) of the full expansion \eqref{eq:potential-generic}. For the T-model, imposing again that $\Re\left(\mu_k^{(T)}\right)>0$ in \eqref{eq:floquet-T} we obtain
\begin{equation}\label{eq:instab-scale-T}
    l_{\text{inst}}^{(T)}=2\Biggl(-(4M^2+6M^2\Phi_1^2+6\lambda\Phi_1^2)\epsilon^2+\sqrt{24M^4\Phi_1^2\epsilon^2+9(M^2-\lambda)^2\Phi_1^4\epsilon^4}\,\Biggr)^{-1/2}.
\end{equation}
Looking at the expression for $\mu_{k}^{(E)}$, eqn. \eqref{eq:floquet-exponent-general}, one can observe the difficulty in obtaining an analytical expression for the instability scale in this case. For that reason, in Fig.~\ref{fig:compare-E}, $l_{\text{inst}}^{(E)}$ is obtained numerically. White dotted lines represent the evolution of the physical modes, and at the moment those modes cross the instability scale, they start to amplify. These crossing points are different if one uses the full expansion \eqref{eq:potential-generic} or just the quadratic term, pointing out the importance of considering higher-order terms. Moreover, the enhancement of perturbations can be physically explained with the negativity of the coefficients of the expansion, $\lambda_3$ or $\lambda_T$. If we treat them as self-interactions of the field, a negative coefficient implies an attractive force, which induces a negative pressure and makes some modes unstable.
Nonetheless, as the field decays, the cubic and quartic terms of the expansion \eqref{eq:potential-generic} became less dominant and one recovers the usual preheating scenario \cite{Kofman:1997yn,Martin:2019nuw,Martin:2020fgl,del-Corral:2023apl} where the potential is well approximated by a parabola and $\tilde{v}_{\bm k}\sim a^{3/2}$. This can be seen in Fig.~\ref{fig:compare-T}, where for small amplitudes of the field $l_{\text{inst}}^{(Q)}\sim l_{\text{inst}}^{(T)}$. For Fig.~\ref{fig:compare-E}, one should go to smaller amplitudes to have this parallelism. Both of these figures are made using $\alpha=10^{-2.5}$. Lastly, it is important to note that not all perturbations with $\left(\frac{k}{a}\right)^{-1}>l_{\text{inst}}$ undergo amplification. The upper limit is set by the Hubble radius $R_H=H^{-1}$ and the set of modes defined by $R_H>\left(\frac{k}{a}\right)^{-1}>l_{\text{inst}}$ is usually called \textit{instability band} (IB). Figs.~\ref{fig:scales-T} and \ref{fig:scales-E} show the IB (blue shaded area) for a T- and an E-model with $\alpha=10^{-2.5}$, respectively. We see again that, for the T-model, $l_{\text{inst}}^{(T)}$ approaches $l_{\text{inst}}^{(Q)}$ sooner that in the E-model case. This is mainly because the former one is closer to the parabola, given the absence of the asymmetric coefficient $\lambda_3$. If we neglect the higher-order terms in the expansion of the potential then we are not considering the amplification of some modes that enter the IB around the beginning of preheating, indicated by the lower gray dashed lines, which could lead to wrong estimations.


\section{Applications}\label{sec:applications}

Floquet theory offers a significant advantage: it allows us to forget about numerical computations and use the analytical results to estimate how the curvature perturbations are amplified. This approach accelerates computations, which typically imply long computational times associated with the highly oscillatory regime of preheating. Furthermore, the amplification process can lead to intriguing phenomena, as we briefly outline in the following discussion.
\begin{itemize}
    \item \textbf{Primordial Black Holes}: Using general relativity one can study the collapse into a black hole (BH) of a real, minimally coupled, massive scalar field in an asymptotically Einstein-de Sitter spacetime background. In \cite{Goncalves:2000nz,Martin:2019nuw,Barroso:2024cgg} it is shown that density perturbations $\delta\rho_{\bm k}$ can collapse into a BH provided that the time the $k$-mode spends inside the IB exceeds the time it needs to collapse, defined as
    \begin{equation}\label{eq:time-of-collapse}
    \Delta t_{\text{coll}}=\frac{\pi}{H[t_{\text{bc}}(k)]\delta_{\bm k}^{3/2}[t_{\text{bc}}(k)]},
    \end{equation}
    where $t_\text{bc}(k)$ is the time at which each mode enters the instability band and $\delta_{\bm k}=\frac{\delta\rho_{\bm k}}{\rho}$ is the density contrast, being $\rho$ the background energy density and $\delta\rho_{\bm k}$ the density perturbations. In our specific scenario, perturbations amplify faster with decreasing $\alpha$ values. This suggests that PBHs form more rapidly and consequently with smaller masses. However, a critical threshold should exist where PBHs may become overproduced, thereby establishing a lower bound for $\alpha$. This is something we intend to study and publish soon somewhere \cite{draft-alpha-PBH}.
    \item \textbf{Scalar-Induced Gravitational Waves}: At second order in perturbation theory, the scalar perturbations couple to the tensor ones, inducing the production of GWs \cite{Maggiore:2007,Guzzetti:2016,Domenech:2021,Baumann:2007zm,Ananda:2006af}. The resulting differential equation for the tensor perturbations $h_{\bm k}(\eta)$ in cosmic time $\eta$ is given by
    \begin{equation}
        h''_{\mathbf{k}}(\eta)+2\mathcal{H}h'_{\mathbf{k}}(\eta)+k^2h_{\mathbf{k}}(\eta)=\mathcal{S}(\mathbf{k},\eta),
    \end{equation}
    where $\mathcal{H}=aH$ is the conformal Hubble rate and $\mathcal{S}(\mathbf{k},\eta)$ is the so-called source function, which directly depends on the metric scalar perturbation $\Phi_{\bm k}$. This last can be related to the curvature perturbation $\mathcal{R}_{\bm k}$ \cite{del-Corral:2023apl} and thus with $v_{\bm k}$, which again suggests that for decreasing values of $\alpha$ the fractional energy density of GWs, $\Omega_{\text{GW}}(k)$, can be high enough to reach the range of detection of some actual and future GWs detectors \cite{Gehrman:2022imk,Aggarwal:2020olq}. The fractional energy density is defined as \cite{Domenech:2021}
    \begin{equation}
        \Omega_{\text{GW}}(k)\equiv\frac{1}{\rho}\frac{\dd \rho_{\text{GW}}}{\dd\ln k}=\frac{k^2}{12\mathcal{H}^2}\sum_\theta\mathcal{P}_{h,\theta}(k),
    \end{equation}
    where $\rho_{\text{GW}}$ is the energy density of GWs and $\mathcal{P}_{h,\theta}(k)$ the power spectrum of tensor perturbations, with $\theta$ accounting for both polarization states. We will also study this and publish our findings \cite{draft-alpha-GWs}.
\end{itemize}


\section{Conclusions}\label{sec:conclusions}

This work introduces an analytical model to explain the amplification of curvature perturbations at small scales during the post-inflationary preheating phase, based on the phenomenon of self-resonance. We consider both symmetric and asymmetric inflationary potentials, \textit{i.e.}, T and E $\alpha$-attractors models\cite{Kallosh:2013daa,Kallosh:2013hoa,Kaiser:2013sna,Kallosh:2015lwa}. Initially, we solve perturbatively for the inflaton field and expand the inflationary potential up to fourth order. Then, we transform the MS equation into a Hill's one \eqref{eq:hill-eq} (see also \ref{sec:appendix-MS}) and provide an explicit formula \eqref{eq:floquet-exponent-general} for Floquet's exponents of the MS variable. This, so far in the literature, was not given when the potential includes both cubic and quartic self-interactions. In addition, the analytical results allow us to estimate, avoiding time-consuming numerical codes, the magnitude of the amplification of curvature perturbations during preheating, as well as the position of the peak in terms of comoving wavenumber $k$ and the instability scale above which perturbations grow. Traditionally, self-resonance has been studied primarily within the framework of the perturbed equation of motion of the field \cite{Kofman:1997yn,Iarygina:2018kee,Iarygina:2020dwe,Krajewski:2018moi,Lozanov:2017hjm,Jedamzik:2010dq,Martin:2019nuw,Sfakianakis:2018lzf,Ballesteros:2024hhq,Martin:2020fgl,Shafi:2024jig,Mahbub:2023faw,Sang:2020kpd,Zhang:2023hjk,Child:2013ria}. However, our approach is applied to the MS equation, which captures the dynamics of the curvature perturbations. While applicable to any potential that can be expressed as \eqref{eq:potential-generic}, we focus our analysis on $\alpha$-attractor models, given their observational favorability \cite{Planck:2018jri}. Particularly, for $\alpha\ll1$, our model reveals a pronounced self-resonance effect, leading to a rapid and substantial enhancement of the curvature perturbations, specifically for asymmetric potentials such as E-models due to the drift term. This analytical model is compared with a full numerical computation, as can be seen in Fig.~\ref{fig:curvatures}, showing a good agreement, especially concerning the T-model. Also, we stress the importance of avoiding the standard parabola approximation during preheating for some models, specifically during the transition from inflation to preheating, as illustrated in Figs.~\ref{fig:compare-quadratic} and \ref{fig:scales}. Moreover, this investigation opens a window on the potential role of $\alpha$-attractors in the formation of PBHs on small scales as well as in the production of SIGWs and oscillons during preheating. Furthermore, it suggests a possible physical lower bound on the parameter $\alpha$, since as we decrease its value, perturbations amplify more drastically, potentially reaching the non-linear regime (see \ref{sec:appendix}) and thus for a very extremely low value of $\alpha$ this could potentially impact on the production of primordial black holes or even scalar-induced gravitational-waves. Finally, we stress that this mechanism of amplification of perturbations is inherent to preheating in $\alpha$-attractor models and does not require any fine-tuning.


\section*{Acknowledgments}

I express my gratitude to João Marto and K. Sravan Kumar for their suggestions and assistance in the development of this work. I am also grateful for the support of grant UI/BD/151491/2021 from the Portuguese Agency Funda\c{c}\~ao para a Ci\^encia e a Tecnologia. This research was funded by Funda\c{c}\~ao para a Ci\^encia e a Tecnologia grant number UIDB/MAT/00212/2020.


\appendix

\section{Mukhanov-Sasaki equation}\label{sec:appendix-MS}

Considering that the matter content is described by a scalar field $\phi$, the solution to the Einstein equations in a perturbed flat FLRW background metric leads to the Mukhanov-Sasaki (MS) equation in conformal time $\eta$
\begin{equation}\label{eq:MS-tau}
    v''_{\bm k}+\left[k^2-\frac{\left(a\sqrt{\epsilon_H}\right)''}{a\sqrt{\epsilon_H}}\right]v_{\bm k}=0,
\end{equation}
where $v_{\bm k}=a\left[\delta\phi_{\bm k}+\phi'\Phi_{\bm k}/\mathcal{H}\right]$
is the so-called MS variable a combination of the field and metric perturbations, $\delta\phi_{\bm k}$ and $\Phi_{\bm k}$, respectively. See \cite{Mukhanov:1990me,Baumann:2009ds} for details. Here, a prime $'$ denotes derivation with respect to conformal time, $\mathcal{H}=a'/a^2$ is the conformal Hubble rate, and $\epsilon_H=1/2\,(\phi'/\mathcal{H})^2$ is the first slow-roll parameter. We are working in Fourier space. The compact form of this equation is not convenient for our purposes, since it does not allows us to transform it into a Hill equation, as we wil shortly see. Therefore, we rewrite it in cosmic time $t$, related to conformal time by $\dd \eta = \dd t/a$. This gives the MS equation in cosmic time:
\begin{equation}\label{MScosmic}
    \Ddot{v}_{\bm k}+H\dot{v}_{\bm k}+\left[\frac{k^2}{a^2}+\frac{\dd^2 V_{T,E}}{\dd\phi^2}-2H^2+\frac{2\dot{\phi}}{H}\frac{\dd V_{T,E}}{\dd\phi}+\frac{7\dot{\phi}^2}{2}-\frac{\dot{\phi}^4}{2H^2}\right]v_{\bm k}=0,
\end{equation}
where the equation of motion of the field, eqn.~\eqref{KG}, has been used. Now, as we did with \eqref{KG}, we will use the time variable $\tau=\omega t$. Also, using the perturbative solution found in \eqref{eq:final-form}, and the expansion of the $\alpha$-attractor potentials, eqns.~\eqref{eq:expansion-T} and \eqref{eq:expanson-E} into \eqref{MScosmic} we obtain, up to order $\mathcal{O}(\beta^2)$
\begin{equation}\label{eq:hill-eq2}
    \frac{\dd^2\tilde{v}_{\bm k}}{\dd\tau^2} +\left(A_k+2q_1\cos(z)+2p_1\sin(z)+2q_2\cos(2z)+2p_2\sin(2z)+h.h.\right)\tilde{v}_{\bm k}=0,
\end{equation}
where we have used trigonometric power-reduction identities to express powers of sines and cosines as sums of harmonics. The term $h.h.$ denotes higher harmonics, and the coefficients are given by:
\begin{subequations}
    \begin{equation}
        A_k=\frac{k^2}{a^2M^2}+1+\left(1+\left(\frac32+\frac{3\lambda}{2M^2}-\frac{\lambda_3^2}{M^4}\right)\Phi_1^2\right)\beta^2,
    \end{equation}
    \begin{equation}
        q_1=\frac{\lambda_3\Phi_1\beta}{M^2},
    \end{equation}
    \begin{equation}
        p_1=-\left(\sqrt{\frac32}-\sqrt{\frac{25}{6}}\right)\frac{\lambda_3\Phi_1^2\beta^2}{2M^2},
    \end{equation}
    \begin{equation}
        q_2=\left(\frac{3\lambda}{2M^2}+\frac{\lambda_3^2}{3M^4}-\frac32\right)\frac{\Phi_1^2\beta^2}2,
    \end{equation}
    \begin{equation}
        p_2=-\sqrt{\frac32}\Phi_1\beta.
    \end{equation}
\end{subequations}
To derive \eqref{eq:hill-eq} we have used that the Hubble rate is given up to $\mathcal{O}(\beta^2)$ by
\begin{equation}
    H^2=\frac13\left(\frac{\dot\phi^2}2+V\right)=\frac{\beta^2M^2\Phi_1^2}6.
\end{equation}
Additionally, to eliminate the damping term, we rescale the MS variable a $\tilde v_{\bm k}=a^{1/2}v_{\bm k}$ \cite{Finelli:1998bu,Martin:2019nuw,Martin:2020fgl}. As it is eviden from the shape of Eqn.~\eqref{eq:hill-eq2}, writting the MS equation in cosmic time (or the rescaled time $\tau$) allows us to transform it into a Hill equation, make use of the perturbative expansion \eqref{eq:final-form} and thus study the amplification of perturbations within the context of Floquet Theory.

\section{Backreaction and non-linearity}\label{sec:appendix}

\begin{figure}
    \centering
    \includegraphics[width=0.5\linewidth]{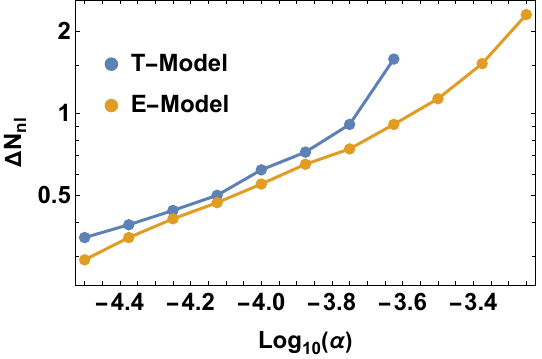}
    \caption{Number of e-folds, $\Delta N_{nl}$, until non-linearity (backreaction) regime is reached as a function of the parameter $\alpha$ and for both T- and E-models.}
    \label{fig:non-linearity}
\end{figure}

The $\alpha$-attractor models have been studied in the context of parametric resonance. See, for instance, \cite{Iarygina:2018kee,Iarygina:2020dwe,Krajewski:2018moi,Lozanov:2017hjm}. Typically, these works employ lattice methods to compute the growth of scalar field perturbations, $\delta\phi_{\bm k}$. Once these perturbations reach $\mathcal{O}(1)$, they start to backreact on the field evolution, leading to its rapid decay. At this stage, linear analysis becames not valid and some interesting non-linear structures may form, such as oscillons or transients \cite{Lozanov:2017hjm}. However, a common limitation of lattice methods is that they often neglect metric fluctuations, $\Phi_{\bm k}$, which can influence the overall evolution of the perturbations.

This work improves earlier estimations by considering the evolution of the MS variable $v_{\bm k}$, which contains both metric and field fluctuations, and is related to the curvature perturbation $\mathcal{R}_{\bm k}$ (see Eqn.~\eqref{eq:curvature-definition}). For comparison, Fig.~\ref{fig:non-linearity} shows the number of e-folds, $\Delta N_{nl}$, until non-linearity (backreaction) is reached for both T- and E-models and different values of $\alpha$, computed using Floquet theory. The evolution is tracked for 5 e-folds after the end of inflation and the results in Fig.~\ref{fig:non-linearity} demonstrate that the non-linearity regime is reached within $\mathcal{O}(1)$ e-fold, except for $\alpha>10^{-3}$. For such values, while self-resonance is still significant, perturbations do not grow sufficiently to reach non-linearity. Conversely, for smaller values of $\alpha$, the non-linearity regime is reached more quickly, within $\mathcal{O}(1)$ e-fold. Although this estimation is similar to the one provided in \cite{Lozanov:2017hjm}, around Eqns.~10-14, our analysis makes the distinction between the T- and E-models clearer, as it considers a polynomial expansion of the potential instead of a monomial or parabolic approach (see for instance eqns.~2-4 of \cite{Lozanov:2017hjm}). In this sense, Fig.~\ref{fig:non-linearity} shows that, due to the asymmetry in the E-model, perturbations amplify faster, reaching the non-linearity regime earlier. Nonetheless, for $\alpha<10^{-4}$, both models exhibit similar behavior.

To further validate the analytical model described in this work, we performed a series of lattice simulations. Due to its ease of use, we choose to work with the publicly available \CL code \cite{Figueroa:2020rrl,Figueroa:2021yhd}. In all simulations, we used a lattice size of $N=128$, with an infrared cut-off of $\tilde{k}_{IR}=0.05$, and parallelized computations across 8 cores. To facilitate the numerical simulations, we introduced the following dimensionless variables:
\begin{equation}
    \tilde{\phi}=\frac{\phi}{M_{P}} \qquad \tilde{k}=\frac{k}{M} \qquad \dd \tilde{t}=M\,\dd t \qquad \dd\tilde{x}^i=M\,\dd x^i,
\end{equation}
where $M_{P}$ is the Planck mass and $M$ the mass of the scalar field. For further details, refer to the \CL documentation \cite{Figueroa:2020rrl, Figueroa:2021yhd}.

\begin{figure}
     \centering
     \begin{subfigure}[b]{0.49\textwidth}
         \centering         \includegraphics[width=\textwidth]{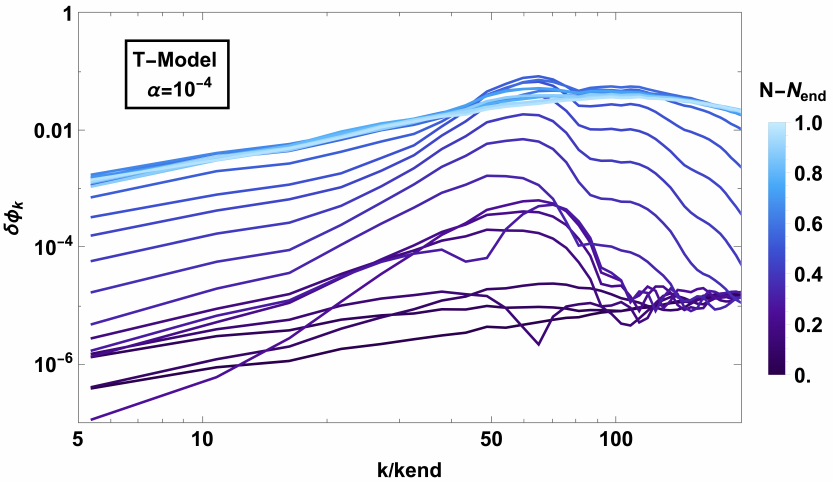}
         \caption{}
         \label{fig:lattice-T}
     \end{subfigure}
     \hfill
     \begin{subfigure}[b]{0.49\textwidth}
         \centering         \includegraphics[width=\textwidth]{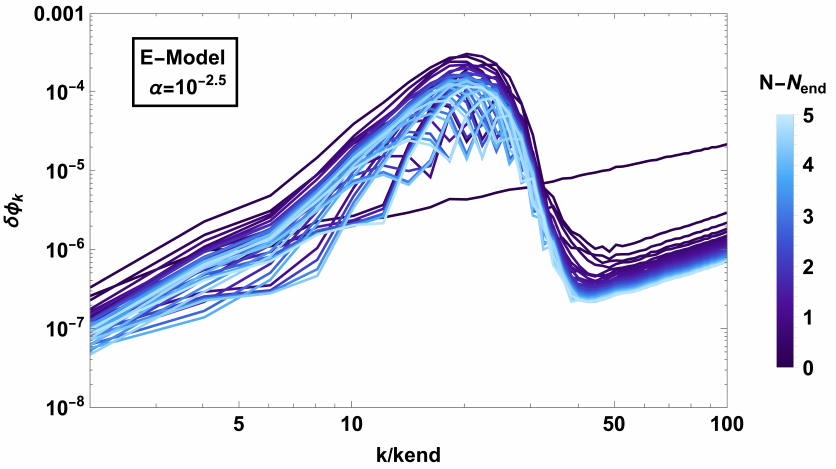}
         \caption{}
         \label{fig:lattice-E}
     \end{subfigure}
     \caption{Lattice simulations during preheating for \textbf{(a)} a T-model with $\alpha=10^{-4}$ and \textbf{(b)} an E-model with $\alpha=10^{-2.5}$. See text for specifications.}
     \label{fig:lattice}
\end{figure}

Fig.~\ref{fig:lattice} shows the computation of the field perturbations $\delta\phi_k$ for two models, a T-model with $\alpha=10^{-4}$ in Fig.~\ref{fig:lattice-T}, and a E-Model with $\alpha=10^{-2.5}$ in Fig.~\ref{fig:lattice-E}. For the former one, since $\alpha<10^{-3}$, we see that at $N-N_{\text{end}}\sim0.5$ the non-linearity regime is reached since the spectrum of perturbations loses its peaked shape and smooths, which confirms our results from Fig.~\ref{fig:non-linearity}. This is explained as follows. Just after the end of inflation, the resonance is strong and around a specific scale. However, as time evolves, and due to the backreaction, the modes start to interact with each other, leading to a redistribution of the energy across different scales, which flattens the spectrum. For the E-model case, where $\alpha>10^{-3}$, the non-linearity regime is not reached, consistent again with our predictions. This is evident from the fact that the perturbation spectrum retains its peaked structure as it evolves. This figure resembles Fig.~\ref{fig:curvature-E25}, since the perturbations in both cases reach similar amplifications and around the same wavenumber $k/k_{\text{end}}\sim 10-20$. Finally, Fig.~\ref{fig:lattice} suggest that the analytical method developed in this work is in good agreement with the numerical lattice simulations.

\bibliographystyle{utphys.bst}
\bibliography{BIBLIO.bib}

\end{document}